\documentclass[10pt,journal,final,a4paper,twoside,compsoc]{IEEEtran}
\usepackage[T1]{fontenc}
\usepackage[utf8]{inputenc}
\usepackage{amssymb}
\usepackage{cite}
\usepackage{flushend}
\usepackage{fontawesome}
\usepackage{graphicx}
\graphicspath{{../img/}}
\usepackage[hidelinks]{hyperref}
\usepackage{textcomp}
\usepackage{xcolor}

\definecolor{insightcol}{rgb}{0.6,0,0}
\definecolor{conjecturecol}{rgb}{0.2,0.0,0.0}
\definecolor{quotecol}{rgb}{0,0.6,0.2}
\definecolor{nonquotecol}{rgb}{0,0.4,0.1}
\definecolor{respondentcol}{rgb}{0,0,0.7}
\definecolor{ivcol}{rgb}{0,0,0.7}

\newcounter{insight}
\renewcommand{\theinsight}{\arabic{insight}}
\newcounter{conjecture}
\renewcommand{\theconjecture}{\arabic{conjecture}}
\newcounter{evidence}[insight]
\renewcommand{\theevidence}{\theinsight\alph{evidence}}

\newcommand{\IV}[1]{\textcolor{ivcol}{{``#1''}}}
\newcommand{\R}[1]{\textcolor{respondentcol}{\textsuperscript{R#1}}}
\newcommand{\rQuote}[2]{\textcolor{quotecol}{``#2''}\R{#1}\index{R#1}}

\newcommand{\Genquote}[1]{\textcolor{nonquotecol}{``#1''}%
\textcolor{respondentcol}{\textsuperscript{Several}}}
\newcommand{\Observations}{\textsc{Observations:} }
\newcommand{\Insight}[2]{\refstepcounter{insight}\label{#1}
  \vspace{1ex}\par\noindent
  \bgroup \color{insightcol}\bfseries Insight \theinsight: #2 \egroup}
\newcommand{\Conjecture}[2]{\refstepcounter{conjecture}\label{#1}
  \vspace{1ex}\par\noindent
  \bgroup \color{conjecturecol}\bfseries Conjecture \theconjecture:
  \color{insightcol} #2 \egroup}

\newcommand{\srcI}{\faUser}  % source: Interviews  Microphone,Comment,User
\newcommand{\srcL}{\faFileText}  % source: Literature  Book,FileText
\newcommand{\Evidence}[2]{\refstepcounter{evidence}\label{#1}
  \par #2 \textsc{Evidence \theevidence:} }
\newcommand{\EvidenceI}[1]{\Evidence{#1}{\srcI}}
\newcommand{\EvidenceL}[1]{\Evidence{#1}{\srcL}}
\newcommand{\EvidenceLit}[1]{\refstepcounter{insight}\Evidence{#1}{\srcL}}
\newcommand{\EndOfEvidence}{\hfill$\square$}

\newcommand{\Heading}[1]{\textbf{#1}}
\newcommand{\Concept}[1]{{\textsl{#1}}}
\newcommand{\Practitioner}{\textcolor{black}{\footnote{practitioner 
  industrial respondent}}}

\newcommand{\citep}[1]{\cite{#1}}

\begin{document}

\title{Four presumed gaps in the\\
       software engineering research community's knowledge
      }

\author{\IEEEauthorblockN{Lutz Prechelt}
  \IEEEauthorblockA{\textit{Freie Universit\"at Berlin} \\
  Berlin, Germany \\
  prechelt@inf.fu-berlin.de}
}

\maketitle

\begin{abstract}
\emph{Background:} 
The state of the art in software engineering consists of a myriad of
contributions and the gaps between them; it is difficult to characterize.\\
\emph{Questions:}
In order to help understanding the state of the art, can we identify
gaps in our knowledge that are at a very general, widely relevant level?
Which research directions do these gaps suggest?\\
\emph{Method:}
54 expert interviews with senior members of the ICSE community,
evaluated qualitatively using elements of Grounded Theory Methodology.\\
\emph{Results:}
Our understanding of complexity, of good-enoughness, and of developers' 
strengths is underdeveloped.
Some other relevant factors' relevance is apparently not clear.
Software engineering is not yet an evidence-based discipline.\\
\emph{Conclusion:}
More software engineering research should concern itself 
with emergence phenomena, 
with how engineering tradeoffs are made,
with the assumptions underlying research works, and
with creating certain taxonomies.
Such work would also allow software engineering to become 
more evidence-based.
\end{abstract}

% \begin{keyword}
%   D.2.0, D.2.19.a, D.2.m, K.6.3.a, K.7
%   assumptions, complexity,
%   developer capabilities, domain, 
%   emergence, evidence-based software engineering, expert interview,
%   good-enoughness, grounded theory methodology,
%   literature survey,
%   software quality, 
%   taxonomy, tradeoffs
% \end{keyword}

%\tableofcontents

%========================================================================
%========================================================================
\section{Introduction}

\subsection{The state of knowledge about software engineering}\label{stateofknowledge}

What do we\footnote{``we'' here is the software engineering research community;
  the people whose job it is to build and explicate knowledge about software
  engineering.}
know about software engineering (SE)?
How does it work?
What is the state of the art?

These questions are difficult to answer, because the things we know
are so specialized that the overall landscape of our knowledge is
highly fragmented.

For instance the 
\emph{Guide to the Software Engineering Body of Knowledge} \citep{SWEBOK14},
despite its common abbreviation as SWEBOK, is only a guide to the
knowledge (a catalog of sources), not the knowledge itself, 
and still comprises more than 300 pages.
It captures well-established knowledge, not the most recent state of the art.

At the SE research front, Kitchenham and others in 2004 suggested
to perform systematic literature reviews (SLRs) in order to synthesize
and consolidate research results in order to enable
evidence-based SE practice \citep{KitDybJor04,Kitchenham04,KitCha07}.
This was successful and many SLRs have appeared since\footnote{There
  are even some SLRs of SLRs (``tertiary studies''),
  but most of these are concerned with research methods
  \citep{KitPreBud10,SilMedSoa11,CruDyb11,ImtBanIkr13},
  not SE methods \citep{HanSmiMoe11}.},
but they help only gradually in fighting the fragmentation,
for several reasons:
Most of them are mapping studies (that merely catalog results)
rather than full SLRs (that synthesize combined results;
see \cite[Table 2]{SilMedSoa11} and \cite[Table 2]{KitPreBud10}),
different SLRs are dominated by different study methods
with different validity characteristics 
(e.g. experiments \cite{HanDybAri09} vs. 
surveys \cite{BeeBadHal08}), and
SLR quality varies widely 
(e.g. compare \cite{KamDybHan07} to \cite{HasHal08}).
Most importantly, however,
few SLRs are about widely applicable areas within SE
(e.g. agile methods \citep{DybDin08},
 motivation \citep{BeeBadHal08})
while most are rather more specialized in terms of their usage contexts;
for instance: agile practices in distributed SE \citep{JalWoh10},
the role of trust in agile methods \citep{HasHal08},
effectiveness of pair programming \citep{HanDybAri09},
search-based SE techniques involving user interaction \citep{RamRomSim19},
unit testing for BPEL routines \citep{ZakAtaGha09}.

This fragmentation is problematic because it makes it difficult to judge
the relevance of a proposed research question: 
What importance does an answer have for SE overall?
How much of an impediment is the gap that the answer might fill?
Sure, we can argue why a given research question A appears
more relevant than question B if A and B are related.
But typically we cannot -- and there is hardly any basis for systematically
\emph{deriving} relevant research questions.

\subsection{What do we not know?}

An alternative approach to making statements about the state of
SE knowledge is to characterize things we do \emph{not} know.
And if we do so 
not at a specialized level (where the list would be nearly endless) but
at a general one, this would overcome the fragmentation.
In that case, each knowledge gap that is pointed out
would have relevance for many, perhaps even most SE situations.

Finding such general-level gaps in the SE knowledge
is the aim of the present work.
The underlying study was largely emergent (rather than designed upfront)
and came to be as follows:

\subsection{A conundrum}

Ever since I entered empirical software engineering research in 1995,
I have been troubled by an apparent paradox:
On the one hand, even modest software is highly complex, few (if any) 
software systems ever work fully satisfactorily, and the people building
those systems all have strong cognitive and other limitations.
On the other hand, despite many deficiencies and local breakdowns,
overall the software ecosystem appears to work OK.
\textbf{How do we}\footnote{Subsequently in this article, ``we'' can stand for 
  the software engineering community (as in the present case),
  the software engineering research community,
  the ICSE community, or
  the union of authors and readers of this article.}
\textbf{manage to achieve that?}

When this question hit me once again the week before ICSE 2018
(International Conference on Software Engineering),
I recognized that asking my colleagues about it might be a source
of insight. 
I worked out a concise formulation and set out to run
a number of interviews during the ICSE conference week.
The rich answers I got prompted me to perform a thorough evaluation, 
which turned into the present article.
I did \emph{not} initially have an idea whether or how my interviews might 
turn into a research contribution; this emerged only during analysis.
Note also that the outcome is about knowledge \emph{gaps}, 
it does \emph{not} answer the above question.

\subsection{Research contributions}

Formulating research \emph{questions} for the present work would be misleading,
as they were
completely emergent; so I formulate contributions only.

Imagine, purely conceptually, a
Software Engineering Topic Tree (SETT)\footnote{This concept will be used
  again in Related Work.}
that arranges SE research articles according to their degree of specialization,
with ``all of SE'' at the top.
Then most SE articles would be at or near leaves,
systematic literature reviews would be at medium levels,
and the present article would be near the top.
In this sense,
\begin{itemize}
\item 
the article derives six areas in which our knowledge is weakly developed
and that each have relevance for a broad variety of SE contexts;
\item 
the article suggests research directions to fill these knowledge gaps.
\end{itemize}
The interview respondents' views are opinions, not facts.
But the gaps in understanding that show up in these views are arguably facts,
not opinions;
the article presents evidence for each of them.
In contrast, the research directions are merely sketched; 
a fuller discussion is left to future work.

%========================================================================
%========================================================================
\section{Method}

I ran 54 semi-structured expert interviews based on a 4-item stimulus card 
and analyzed them using 
various elements of the Grounded Theory Methodology (GTM), 
in particular Open Coding, 
paying particular attention to elements that are \emph{missing} in the
responses.
I report the outcomes in a topic-by-topic manner.

\subsection{Epistemological stance}

My stance is constructionist-interpretive \citep{Charmaz08}, not positivist:
We should not assume the existence of a single underlying truth;
the data requires interpretation and the person of the researcher
will unavoidably have influence on the outcome;
alternative interpretations of these same data may exist
that have similar validity.
This does not threaten the \emph{validity} of the present interpretation,
but in terms of \emph{appropriateness}, individual readers may prefer
alternative interpretations in some places.

\subsection{Design goals and approach}

Although the study overall does not have an upfront design,
the interviews have.
I pursued the following ideas:
\begin{enumerate}
\item \emph{Openness:}
  Due to the lack of a better idea, the study would use
  my own perception of the situation as a starting point
  (as sketched in the introduction),
  but beyond that I avoided to direct the
  attention of my respondents in any particular direction.
  The approach was to use basically just a single question.
  As far as I can see, this has worked well.
  For the same reason, I kept interjections and followup questions during the
  interviews to a minimum.
  This has worked well for many interviews, but presumably made some others
  poorer.
\item \emph{Breadth over depth:}
  I expected responses to be highly
  diverse (which indeed they were) and thus preferred getting many short
  interviews over fewer long ones.
\item \emph{Respondent diversity:}
  I strove to represent women
  disproportionately highly, because I suspected their views might have a
  different distribution.
  I strove to have respondents from as many countries as I could,
  because I suspected that national perspectives might differ.
  With one exception, a PhD student, all respondents are senior researchers.
  This may or may not have been a good idea.
\item \emph{Provocativeness:}
  To reduce the risk for boilerplate-style
  responses, the interview stimulus used
  pointed formulations that I expected many respondents to disagree with.
  The formulations were also intentionally ambiguous to prompt differentiation
  (an important aspect, please keep it in mind).
  Both of these approaches worked in many if not all cases.
\end{enumerate}

\subsection{Interview structure: The stimulus card}

\begin{figure}
  \includegraphics[width=\columnwidth]{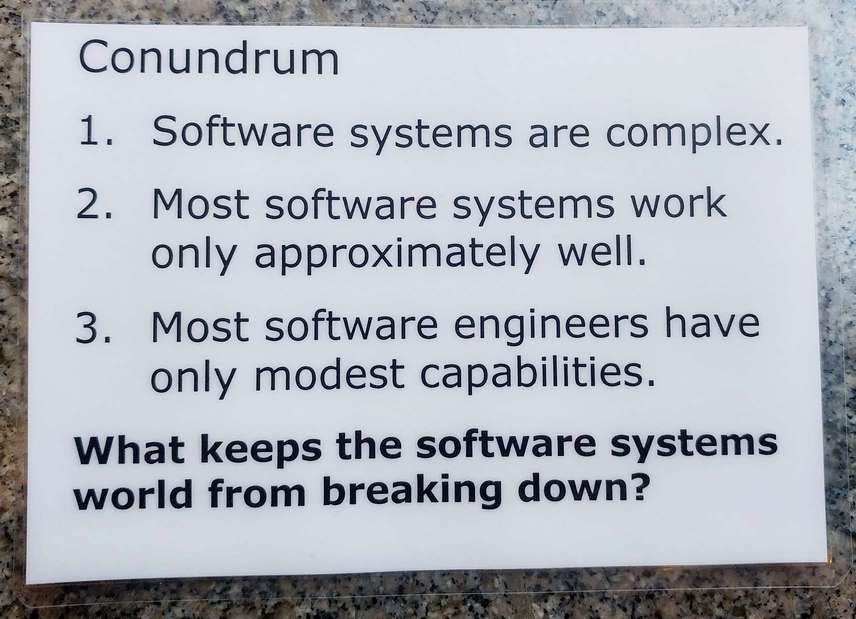}
  \caption{Stimulus card (front)}\label{stimulus.jpg}
\end{figure}

The interview was based on the 142~mm by 104~mm laminated card shown
in Figure~\ref{stimulus.jpg}.

Its rear side contained a minimal informed consent agreement
appropriate for researchers:
\begin{quote} \small\sffamily
May I record this? (audio)\\
Anonymous.\\
Will quote only anonymously.
\end{quote}
All respondents agreed to this. 
One requested to have the recording deleted after transcription, which I did.

The front side contained three statements (which I will call S1, S2,
S3) meant to establish a context for the main interview
question (which I will call QQ).
\begin{itemize}
\item \textbf{S1}: Software systems are complex.
\item \textbf{S2}: Most software systems work only approximately well.
\item \textbf{S3}: Most software engineers have only modest capabilities.
\item \textbf{QQ}: What keeps the software systems world from breaking down?
\end{itemize}

\subsection{Interviewer behavior}

To acquire respondents, I would approach a candidate physically and
either ask
\IV{Hi <firstname>! Would you have time for a 5-minute, 1-question
interview?}
or: ask
\IV{Hi <firstname>! Would you have time for a 5-minute interview on this
question?} and hand them the stimulus card for review.

Most interviews started there and then.
Others were postponed one or more times because the candidates were
busy or hesitant.
In some cases this did not converge and they are now missing.

At interview start, I would almost always hand over the stimulus card so
that the statements and question became physically tangible and could
be pointed to. My only prompt would be something like \IV{OK, shoot.}.

During the interview, my only common interventions would be these:
\begin{itemize}
\item When respondents spoke about individual software systems rather
  than the whole ecosystem, I would emphasize the question was about the
  \IV{software systems \emph{world}} and then often elaborate that this
  meant \IV{widespread collapse}.
\item When respondents gave rather general answers, I would ask for
  specific \IV{ingredients}, \IV{aspects}, or \IV{factors}
  that were most relevant for how we prevent breakdown.
\item Once one or more key ingredients had been named, I would often
  ask the follow-up question
  \IV{Are ICSE and the contributions it is seeing well aligned with
    what's important for bringing software engineering forward?}.
  Its actual formulation 
  often picked up statements the respondent had made
  before.\footnote{The answers to this question will not be discussed here,
  but rather in a separate arXiv article.} % \cite{Prechelt19-icsealign}.
\item I sometimes asked short questions for clarification or
  repeated, in different words, what I had understood.
\end{itemize}
I generally avoided nudging respondents in any particular direction.
The interview ended when the respondent had nothing more to say.

Later interviewees overheard parts of an interview in about
10\% of the cases.

\subsection{Equipment used}

The only equipment used for the interviews were the laminated 
stimulus card and an Olympus VN-8700PC mono voice recorder,
which produced clear recordings despite the often enormous background noise.
Both fit easily in my jeans pockets so that I was ready for an interview
at all times.

\subsection{Respondent demographics}

Of the 54 respondents, 18 (33\%) were female.
The respondents came from 18 different countries of affiliation:
USA (24 interviewees), Germany (5), Canada (4), UK (4), 
Switzerland (2), India (2), Netherlands (2), 
Australia, Spain, Finland, Greece, Ireland, Israel, Italy, 
Luxembourg, South Africa, Sweden, Singapore.

All interviews but one were held in English; 
I noticed two respondents having some difficulty expressing their thoughts 
in English, all others appeared fluent.

16 respondents have previously been chairs of program committees
of the ICSE Technical Research track,
another 5 have been ICSE General Chairs, and
yet another 8 have been PC chairs of other ICSE article
tracks such as SEIP, NIER, SEIS, SEET.
So at least 29 respondents (54\%) would be considered \emph{very}
senior and many of the others were similarly accomplished.

\subsection{Interview metadata}

I ran 51 interviews during the ICSE week (from Sunday 2018-05-27 to 
Friday 2018-06-01)
and 3 more within seventeen days after that, with people who could not 
make time for the interview during the ICSE week; 54 interviews overall.
These range from 2 to 34 minutes in length, 
the middle half being 4 to 10 minutes.
A few interviews consist of two parts, where a respondent approached me
again shortly after the interview to add something.

Respondent quotes will be attributed to respondent pseudonyms 
chosen according to the names of the recording files: 
\R{394} to \R{488} (with gaps).

\subsection{Data analysis}

Analysis used elements of Grounded Theory Methodology GTM 
\citep{StrCor90,Glaser78,Charmaz06}.
The approach is eclectic, taking the basic concepts of
Open Coding \cite[II.5]{StrCor90},
Constant Comparison \cite[II.6]{StrCor90},
Selective Coding \cite[II.8]{StrCor90}, applied multiple times,
the ``as much as needed'' approach to transcription \cite[p.30]{StrCor90},
and some Memoing \cite[III.12]{StrCor90},
from Strauss and Corbin,
a particular emphasis of Theoretical Sensitivity from 
Glaser \cite{Glaser78} and Strauss/Corbin \cite[I.3]{StrCor90} alike,
and the Constructionist epistemological stance
(with some resulting aspects such as insistence on recordings as opposed to 
mere notes and heavy reliance on quotations in the present report)
from Charmaz \cite{Charmaz06,Charmaz08}.

The outcome does not claim to be a theory, so it is unproblematic
that other GTM elements are missing:
Axial Coding \cite[II.7]{StrCor90} is hardly applicable to this data,
Theoretical Sampling \cite[II.11]{StrCor90}
and determining Theoretical Saturation \cite[p.178]{StrCor90}
were impractical.

The analysis started by turning the recordings into text.
This served two purposes: Speeding up subsequent steps and protecting
respondent anonymity in the 
published data (see below).
I transcribed concise pertinent parts verbatim
(shown like this: \rQuote{123}{verbatim quotation}, if it came
 from respondent \R{123}),
paraphrased verbose pertinent parts near-verbatim
(\rQuote{123}{[paraphrased quotation]}),
and paraphrased less-and-less relevant parts with higher-and-higher compression.
In some cases I subsequently expanded paraphrased parts.

Right after textifying each interview, I annotated them with a preliminary 
set of concepts that grew in Open Coding manner over time,
using the MaxQDA 2018\footnote{\url{https://www.maxqda.com/}} software.

In a second pass through the material I created a very short (1 to 6
lines) summary of each interview, all in a single file, to
capture the gist of each interview in order to avoid distortion from
over-emphasizing low-relevance aspects.

During this second pass, I recognized the three different frames of
reference relevant in these data (Section~\ref{nature}) and
discovered the core categories that led to the main insights
(Sections \ref{insightfirst} to \ref{insightlast}).
As a consequence, rather than going for a completely new second
generation of concepts for my annotations as planned, I merely consolidated the
first generation enough to support finding the pertinent material
and validating the statements.
I limited the annotation precision to the level needed to support the 
rough quantification provided here (see Section~\ref{quantification}) 
and ensure the validity of the quotations.

All remaining analysis was then driven by the development and
validation of the narrative you find in this article.

\subsection{Quantification}\label{quantification}

Given the opinions presented below,
it is tempting to ask for quantification of their frequency.
We will have to resist that urge.

Not only is it unclear what such frequencies would mean, because
we are not looking at a \emph{representative} sample of our population.
Worse: Although GTM tailors the categories to the data,
the frequencies cannot be determined properly, because
many responses are highly ambiguous or multilayered.\footnote{It is
  possible to nail down categories completely despite such effects,
  but this is \emph{extremely} laborious and would not pay off
  in the present case. 
  For instance \cite{SalPre13-baseconbook} required about 7
  person-years of work to understand 60 concepts precisely.}

To reflect at least large differences in frequency, I will use
a coarse, 4-level, linguistic, roughly logarithmic ordinal scale as follows:
``most'' means at least 1-in-2;
``many'' means at least 1-in-4;
``some'' means at least 1-in-8;
``few'' means at least 2, but less than 1-in-8.
The basis for these ratios is always smaller than 54, 
because there is no single aspect that all respondents commented on.
Except where noted explicitly, I do not report concepts that occur only once
in my data.

\subsection{Literature micro-studies}\label{microstudies}

To roughly triangulate \citep{Yin03}
the interview-based evidence with a very different 
data source, I performed a number of micro-studies.\footnote{Another one
  is for supporting a statement in the Related Work section.}
Most of these consider the 153 articles from the ICSE 2018 Technical Research 
track (including the 47 Journal First entries) in order to estimate, 
roughly, how often SE research addresses a certain issue.
The classification is made based on article title, session title, and
(where needed) the abstract.
The outcomes serve as modest corroboration of the interview
interpretations, not as research results in their own right.
Details for the micro-studies are included in the above-mentioned
raw data package as well.

\subsection{Member check}

I sent a complete draft of the article to the respondents for a member check,
asking for possible objections to 
each person's quotations,
the insights, and 
the conclusions.
Ten respondents reacted.
As a result, I corrected the handling of two quotations and inserted
minor clarifications in a few places.
There were some comments on, but no objections against any of the insights
or conclusions.

\subsection{Chain of evidence}\label{chainofevidence}

The chain of evidence for the present study consists of
(1)~the raw interview data,
(2)~a set of first-generation and
(3)~second concepts from open coding,
(4)~concepts from selective coding, and
(5)~the micro-studies.
In this article,
(1)~is represented by loads of verbatim quotations,
(4) and (5) as argumentative text.
(3)~is mostly and (2) is completely suppressed for sake of readability;
see Section~\ref{limitations} for a discussion.

A raw data package (interview transcripts, annotations, code memos, etc.)
for this study is available to fill in details.\footnote{%
\url{https://www.dropbox.com/s/woabq7kuxipojz8/knowgaps-materials.zip?dl=0}}
Quotations reported in the article were chosen for their brevity first
and to show the breadth of opinions and formulations second.

%========================================================================
%========================================================================
\section{Deduction Method and Nature of the Results}
\label{deduction}\label{nature}

When you think about the results below,
be aware there are three different frames of reference involved:
\begin{itemize}
\item Software engineering space (SES) is the space of real-world
  SE phenomena.
\item SE knowledge space (KS) is the space of what
  we, the SE research community, know about SES.
\item Interview space (IS) is the space of what my respondents say
  about SES and KS.
\end{itemize}

When interpreted as statements of fact at the SES level, 
the respondents' IS statements can be dismissed as
mere opinions, but this is \emph{not} the point of this article and
not its level of discussion.
Rather, the article mainly uses the IS statements to draw conclusions about KS
--- and mostly does so by analyzing what is \emph{not} being said
with respect to S1, S2, S3, or QQ.

The reasoning works like this:\\
1.~QQ asks for a rough theory of how SE works and S1, S2, S3 are relevant
only to help stating this theory.\\
2.~If none or nearly none of the expert respondents use some concept X
although (i)~X would be useful for their discussion of S1, S2, S3, QQ
and (ii)~the respondents are researchers 
and therefore normally justify their views,
I conclude that knowledge of X is not or not readily available to
those respondents. Let us call this conclusion CX.\\
3.~Since the respondents are many, broadly selected, senior
representatives of the ICSE community,
I conclude that CX holds for the ICSE community as a whole.\\
4.~Since ICSE represents top-quality SE research broadly,
I further conclude that CX holds for the SE research community as a whole.

%========================================================================
%========================================================================
\section{S1: Software Systems are Complex}\label{S1}\label{insightfirst}

If respondents comment on S1, S2, or S3 at all (many did not),
I expect their comments to provide support for their answer to QQ
and analyze them from that perspective.

The result sections on S1, S2, S3, and QQ follow a common pattern:
They first describe general observations about the responses, 
then formulate an insight, and
then present evidence supporting that insight.

% 27 interviews with S1 markup

\Observations 
S1 was meant as a common starting point, 
a self-evident consensus statement present only to 
set the stage for the controversial statements S2 and S3 following it.
And indeed almost everybody of the 27 respondents who commented on S1 
agreed with it:\\
%\rQuote{440}{True} or\\
\rQuote{471}{I definitely agree that's the case.} or\\
\rQuote{462}{And yeah, of course: Software is complex} or\\
\rQuote{452}{I'd agree with all three of those; that seems fairly 
  uncontroversial to me.}
and so on.
A few remarked it wasn't unconditionally true, e.g.\\
\rQuote{445}{I'm willing to accept that as a statement [sic!]; not all software
  systems are complex}.

To my surprise, two respondents rejected S1.
They both referred to the execution state space and argued that
only a tiny part of it was actually relevant:\\
\rQuote{400}{A small number of variables matter} or\\
\rQuote{414}{actually [developers] don't have to reason about all of it, 
  they just have to reason about a little bit of it}.

Reviewing the evidence offered by the large majority that had agreed
to S1 led to 

%****************************************
\Insight{Icplx}{Our notion of complexity is undercomplex.}

I will first explain this by evidence taken from the interviews 
(indicated by the \srcI\ symbol),
then by further evidence from a micro-study
(symbol \srcL).

\EvidenceI{EcplxNodef}
The respondents who agreed with S1 offered no evidence in support of their
claim nor even a definition of complexity.
Where an implicit concept of complexity became visible
at all, it was either size
(\rQuote{452}{you're asking about software systems that are complex, 
  not small ones; [...] a few million lines of code})
or it referred to some effect of complexity
(\rQuote{467}{[...] not just that it's too complex to understand [...]}).

Few respondents differentiated complexity into different
kinds (technical vs. socio-technical; \R{416},\R{446}) or 
complexity from different sources
(components, interaction, algorithm, distribution, scheduling; \R{396}).
No respondent offered a reference that would spell out the discriminations.
\EndOfEvidence

\EvidenceL{EcplxNoemergence}
There is a trans-disciplinary literature on Systems Science and
Complex Systems \citep{MobKal15,wp:complexity}
which offers a number of attributes associated with complexity,
such as nonlinearity, emergence, self-organization, adaptation, 
or feedback loops.
Of these, the one most broadly applicable to software is emergence,
but current SE literature has little discussion of that.
Searching for the word stem ``emerg'' in the PDFs of the ICSE 2018
technical research track finds 144 matches in 35 articles, but
zero of those relate to software complexity.
The closest miss, a work on ``emerging issues'',
talks about a merely temporal aspect \citep{GaoZenLyu18}.
\EndOfEvidence

%========================================================================
%========================================================================
\section{S2: Most Software Systems Work Only Approximately Well}
\label{S2}

% 33 interviews with S2 markup

\Observations
The purposefully ambiguous and provocative S2 question
resulted in a lot more discussion.
Many of the 33 respondents who commented on S2
criticized the formulation, e.g. 
by pointing out that SE shoots at a moving target
(\rQuote{439}{Requirements evolve}).
Two relevant criticisms recurred.
First, some respondents found the word 'only' to be overly negative, e.g.\\
\rQuote{400}{[...] that's a celebration!} or\\
\rQuote{419}{'Approximately' is what you want in engineering}.
% \rQuote{448}{'Only' has negative connotation, but most systems work
%  sufficiently well.}

Second, some pointed out that software is part of a socio-technical 
system. In some cases this was apparently crucial for giving 
a positive answer:\\
\rQuote{414}{Software plus society: yes.} or\\
\rQuote{430}{they are working sufficiently well so that they are useful.} or\\
\rQuote{448}{most systems work
  sufficiently well. They support their community, support society.}

In one case it was reason for a strongly negative one:
\rQuote{446}{I think of what we are doing as racking up social debt. 
  Analogous to technical debt. Big time.}.

Overall, half the respondents who commented on S2 agreed with it;
some disagreed, nearly all of them in the optimistic direction
(\rQuote{440}{Not true [...]. I think they work very well.} or
\rQuote{445}{Most software systems that are used work well for the purposes 
  their users expect of them and therefore they work well.})
and many took no clear position.

But lurking behind all this differentiation is a lot of vagueness
in a relevant area:

%****************************************
\Insight{Igoodenough}{We lack understanding of what makes software
  ``good enough''.}

I will again first explain this by evidence taken from the interviews,
then by further evidence from a micro-study.

\EvidenceI{EgeDef} 
The idea of software being 
\emph{good enough} and working 
\emph{sufficiently well} came up in many responses and in many
guises, e.g.\\
as a self-evident observation:\\
  \rQuote{397}{Approximately well may be well enough, that's the point.};\\
as typical-case satisficing:\\
  \rQuote{397}{they work approximately well. That's often good enough; 
    it's usually good enough.};\\
as a sufficient-for-praise level of quality:\\
  \rQuote{418}{But we have many many computer systems doing amazing things 
    all the time; and working very well, working well enough.};\\
as an investment decision and engineering tradeoff:\\
  \rQuote{396}{It's always how much do you want to invest to get your software
    right and when is it good enough};\\
as defined by socio-technical criteria:\\
  \rQuote{430}{Many systems are working and they are working sufficiently well
    so that they are useful.};\\
as a morally justifiable standard:\\
  \rQuote{448}{In most cases we're building systems that are good enough 
    for their purpose and there's nothing wrong with that.};\\
as a matter of life and death:\\
  \rQuote{471}{I want my airbags to go off; and those work pretty well};

%\rQuote{418}{But approximately well may be good enough.}
%\rQuote{419}{The point is to quantify the 'approximately'}

Like the latter, many comments on S2 and (more often) QQ suggested
that high-stakes software (\Genquote{airplanes}) usually worked well
while low-stakes software (\Genquote{apps}) often did not but that this
was acceptable.
But what about the huge region in between: medium-stakes software?
This was hardly ever mentioned and nobody ever offered a definition
of ``good enough'' for this realm.
Indeed, given the multitude of perspectives offered above, 
a definition is not obvious.
\EndOfEvidence

Given that this notion appears to sit at the very center of software
quality, SE research without such a definition 
does not appear good enough.

\EvidenceL{EgeICSE}
In a literature micro-study, I looked for ICSE 2018 articles that appeared
to have any kind of conscious quality tradeoff as a main concern.
This is considerably wider than good-enoughness and so provides a conservative 
estimate how big the problem formulated by Insight \ref{Igoodenough} is.
Considering all 153 articles, 
% and 20 of the abstracts
the search found only five such articles\footnote{Three of these are 
  journal-first publications.}
\citep{HadHasAya17,HeaSadMur18,RamKem19,RehMirNag18,ScaDiNArd17}, or 3\%.
\EndOfEvidence

%========================================================================
%========================================================================
\section{S3: Most Software Engineers Have Only Modest Capabilities}
\label{S3}

% 36 interviews with S3 markup

\Observations 
Of the 36 respondents who commented on S3, most agreed with it, 
if in very different ways:\\
from
\rQuote{471}{Not all programmers are super-programmers.}\\
over
\rQuote{416}{Sure, I don't know what more to say.}\\
and
\rQuote{415}{we know software engineers are not very good at their jobs.
  I work at one of the best places in the world\Practitioner 
  and still...it scares me.}\\
to
\rQuote{467}{I've seen software developers; they do mostly suck.}.\\
These statements resonate with topics we are all familiar with from
the SE literature when developers are discussed:
they often commit mistakes, often produce deficient designs, are often lazy,
often lack discipline.

Some respondents, however, \emph{dis}agree with S3 -- 
also in very different ways:\\
from
\rQuote{419}{To be a software engineer, you need to have good capabilities.}\\
over
\rQuote{459}{I think most people who develop software and have only modest
  capabilities are not software engineers.}\\
to
\rQuote{421}{we're talking about the most intelligent people on the 
  planet here}.

Other comments state that (1)~capabilities vary widely, 
(2)~needed capabilities vary widely as well 
(\rQuote{399}{People find software systems where the complexity matches their
  capabilities.})
and (3)~a few brilliant people can achieve a lot as 
enablers for the others 
(\rQuote{395}{We get the good people to design the essence, so that
  everybody else can build the peripheral stuff that is not as critical.
  Which means that the people that are average can contribute as much as 
  the people that are great.}).

Which sounds a lot more optimistic, but also curiously unfamiliar from the
research literature. Which leads to

%****************************************
\Insight{Icaps}{We don't know much about what software engineers 
  are good at.}

\noindent This time, we rely on a micro-study only.

\EvidenceL{EcapsCount} 
We determine how many works at ICSE have 
one or more developer strengths as a main concern or
one or more developer weaknesses or
both or neither -- and expect the latter group to be largest.
For instance all works that focus on product rather than process will 
automatically land in the ``neither'' group.
Considering all 153 articles,
% and again 20 abstracts, 
we find 14 studies (9\%)
that are developer-behavior-centric (about actual behavior,
not tasks, roles, practices, or expectations), 
but only six of them (a mere 4\%) are outside the ``neither'' category:
Four concern weaknesses 
\citep{BelSprSpi18,GerRobPoo18,HadHasAya17,KruWieFen18}, 
one concerns strengths  \citep{RehMirNag18}, and 
one concerns both  \citep{FreRasAnt19}.\footnote{Two of these six are
  journal-first publications.}

Apparently, most developer-centric work
is merely descriptive, not evaluative.
Within evaluative works, there indeed appear to be more regarding weaknesses
than strengths, but both types are rare.
\EndOfEvidence

%========================================================================
%========================================================================
\section{QQ: What Keeps the Software Systems World From Breaking Down?}
\label{QQ}\label{insightlast}

% 53 interviews with QQ markup

\Observations
While nobody claimed or assumed \emph{everything} was fine
and many respondents talked about various cases or degrees of local breakdowns,
only few respondents rejected the assumption behind QQ and
stated they expected the software systems world to likely break down
(\rQuote{407}{There's very little from keeping the software systems world from
  breaking down.})
or saw it as breaking down already
(\rQuote{436}{I do not think that the software systems world is 
  not breaking down. It is.}).

At the other end of the spectrum, only few respondents expressed that they
firmly expected the software systems world to \emph{not} break down
(\rQuote{439}{I'm totally impressed how software is being constructed.})

The big majority, however, took the assumption of non-breakdown for 
granted and focused on what they considered key factors preventing breakdown.
A majority would initially state only one key factor
and add others only after prodding.
The ensuing overall list of presumed key factors is long and diverse.
I consolidated the list by grouping related concepts 
(which had originally been formed according to an intuition of 
sufficient differentness) until, 
with two exceptions that appeared too relevant, they all had ``some'' mentions
(that is, in at least 7 interviews) or ``many'' mentions.

There is only one factor with many mentions: 
The \Concept{developers}.\\
\rQuote{431}{Of course, people are always the most important.} or\\
\rQuote{460}{It's because people are creative.} or\\
\rQuote{488}{It isn't our research community that keeps it from breaking down,
  that's for sure. I think it's the people in industry that
  figure out what's happening and adapt.}.

After that, there are 11 factors mentioned in more than a few interviews
(in decreasing order of frequency):
\begin{itemize}
\item An \Concept{appropriate development process}\footnote{``appropriate''
    does not imply well-defined, fully orderly, etc.}:\\
  \rQuote{426}{Disciplined practices} or\\
  \rQuote{403}{a reasonable process} or\\
  \rQuote{462}{we have learned to think about our processes}.
  %\rQuote{412}{The process that everyone in the ecosystem follows.}.
  %\rQuote{467}{They probably need to be told how to care. Like a checklist.}.
\item The \Concept{flexibility of users}:\\
  \rQuote{418}{Systems actually do break down all the time, but 
    as humans we work around these systems.} or\\
  \rQuote{445}{Humans are resilient; the human world is resilient.
    Human processes step in} or \\
  \rQuote{433}{expectation management. [We have learned to tell people 
    this is the best they will get -- with some bugs in it.]} or\\
  \rQuote{434}{People are incredibly adaptive.}.
\item \Concept{Abstraction}:\\
  \rQuote{485}{It all boils down to abstraction.} or\\
  \rQuote{438}{'Module' would probably be my biggest.}.
\item \Concept{Software adaptation/evolution}:\\
  %\rQuote{412}{a commonly agreed-upon [process], which is evolving as we
  %  go along} or\\ % process, not product
  \rQuote{418}{We organically grow the systems to be more and more complex} or\\
  \rQuote{483}{we continue to invent}.
\item \Concept{Quality assurance}:\\
  \rQuote{474}{inspections} or\\
  \rQuote{428}{a lot of testing being absolutely indispensable.}.
  %\rQuote{420}{We made a lot of advances in testing, quality control}
\item \Concept{Repairing flaws}:\\
  \rQuote{403}{[Few of our errors] manifest in visible faults.
    And when they do, we fix them.} or\\
  \rQuote{483}{we continue to fix some of the things that weren't
    working as well in the past.}.
\item \Concept{Good-enoughness} (as a required level of quality far below
  perfection):\\
  \rQuote{436}{Beta-testers find stuff and everybody assumes that's 
    just fine.} or\\
  \rQuote{462}{Most of the time, no really terrible things happen.}.
\item \Concept{System resilience} against imperfections:\\
  \rQuote{462}{Fault-tolerant software} or\\
  \rQuote{486}{Code is less sensitive to input changes than we think.}.
\item Development \Concept{tool support}:\\
  \rQuote{395}{Tooling helps. Much more so today than in the past.} or\\
  \rQuote{420}{a lot of advances in computer-supported cooperative work.}.
\item Developer \Concept{collaboration}:\\
  \rQuote{453}{Teamwork.} or\\
  %\rQuote{412}{It's a collaborative effort.} or
  \rQuote{424}{People together are intelligent enough for fixing stuff that
    doesn't properly work.}.
\item \Concept{Reuse} of software and approaches:\\
  \rQuote{395}{many systems today are built out of components that have 
    proven their worth.} or\\
  \rQuote{460}{Solution by analogy.}.
\end{itemize}

I left two factors as separate concepts although they were mentioned in only 
a few interviews. Both apply predominantly to critical systems:
\begin{itemize}
\item \Concept{effort/investment}:\\
  \rQuote{430}{in many software organizations things get built by brute force.
    Built and maintained by brute force.
    That's not [efficient], a lot of effort is wasted, but especially 
    critical systems like flight software, Airbus, things are verified 
    again and again.} or\\
  \rQuote{425}{A lot of manpower is invested} or\\
  \rQuote{436}{an enormous amount of redundancy and recheck and carefulness
    which is built into all of the processes to make sure that 
    that stuff works. And so they check and recheck and recheck and
    recheck and recheck.}.
\item \Concept{formal methods}:\\
  \rQuote{414}{adopting formal methods that have matured enough to 
    get some incremental value out of it.}.
\end{itemize}

All of these 14 were explained in some way by at least a few respondents
and appeared sensible and sufficiently important to me
to consider them all valid factors.
As a list of ``key'' factors they are already many, but looking at them 
more closely led to

%****************************************
\Insight{Idiv}{There is no consensus on a small set of neat key
  software engineering
  success factors.}\footnote{Note that ``success'' here is just an
    abbreviation for non-breakdown.}

I will first explain how the set is not small,
then how the concepts suggested are not neat.

\EvidenceI{EdivA} 
The above factors list is misleading in the sense that it may suggest each item
is a single, neat concept. 
And indeed that may be sort of true for some, for instance 
\Concept{abstraction} and \Concept{collaboration}.
But most are in fact collections of related, but different concepts.
For instance the biggest one, \Concept{developers}, 
upon closer inspection falls apart
into the remainder represented in the examples given above
and two clusters. 
Cluster 1 talks about developers' attitude:\\
\rQuote{409}{people with a particular sense of quality.} or \\
\rQuote{415}{Maybe I'm a paranoid software engineer\Practitioner 
  and that's the appropriate thing to be.}.\\
Cluster 2 talks about high developer capabilities:\\
\rQuote{394}{exceptional people} or\\
\rQuote{395}{We get the good people to design the essence, so that 
  everybody else can build the peripheral stuff that is not as critical.
  Which means that the people that are average can contribute as much as
  the people that are great.}.\\
Cluster 2 \emph{further} falls apart into comments regarding the crucial role of
the top-talented developers
(as in the quotations above)
and other comments assuming that talent is \emph{generally} high
(as in the \R{488} quotation at the top).
Likewise for most other concepts.
\EndOfEvidence

\EvidenceI{EdivB} 
Futhermore, it is nearly impossible to make these concepts orthogonal and
independent of each other.
For instance 
the concept behind the \R{395} \Concept{developer} quotation above
is intertwined with the \Concept{abstraction} concept;
\Concept{good-enoughness} can only be understood via
\Concept{users' flexibility};
the \Concept{effort/investment} talks about a style of
\Concept{quality assurance};
and so on.
\EndOfEvidence

So now we know that perceived SE success factors are diverse.
But what about the validity of the specific factors offered by the respondents?
Investigating the evidence offered by the respondents led to

%****************************************
\Insight{Iebse}{Software engineering is so far not an
  evidence-based discipline.}

I will make three arguments based on the kinds of evidence offered in
the interviews\footnote{Please keep in mind Section~\ref{deduction}.}
and a fourth based on a comparison with medicine.

\EvidenceI{EebseA} 
Many respondents did not offer any evidence at all.
This occurred in two forms: 
In form 1, the factor would simply be stated, period.
Form 2 was less obvious:
Mentioned factors were often accompanied by examples of all kinds,
from vague hints to phenomena through to specific titles of research articles.
% 5 S1, 12 S2, 2 S3, 46 QQ = 65
There were 65 such examples overall throughout the study, 
many of those not single examples but a short salvo of two to four related ones.
The roles of these examples where split about half-and-half into
corroboration of a statement on the one hand 
(see Evidence \ref{EebseB} and \ref{EebseC} for a discussion of these)
versus mere clarification on the other\footnote{I tried to find quotations 
  to illustrate the difference, but they all would require so much context
  from the interview that they are not practical to present.}
-- the latter not constituting even an attempt at providing 
evidence. 
\EndOfEvidence

\EvidenceI{EebseB} 
In some cases, the example offered was highly underspecific,
coming from a non-software domain, which suggests SE evidence
is less readily available:
\rQuote{416}{And when those edge cases happen is when we find things 
  collapsing. The Shuttle breaking apart}.
Here, ``edge cases'' relates to software logic but the example given does not.
\EndOfEvidence

\EvidenceI{EebseC} 
Most evidence offered was informal, 
most often referring to certain categories of software, 
less often to specific software systems, some non-software domain,
personal experience, or 2018 ICSE keynote talks.
Respondents made only 11 references to specific research works 
and only about
half of those were used to corroborate a claimed success factor.
\EndOfEvidence

\EvidenceL{EebseD} 
Contrast this with the situation in a more strongly evidence-oriented 
discipline:
In medicine, there is the Cochrane Collaboration, 
an open consortium of researchers from over 130 countries
who prepare systematic reviews and meta-analyses of the evidence
provided in the medical literature.
On their website \cite{CochraneDiabetes}, 
one can find 391 such reviews for the search
term ``diabetes'' alone; 
works that meta-evaluate treatments for diabetes
or diabetes as a complicating factor in the treatment of other diseases.
A medical researcher would presumably refer to some of this evidence
in a context where it is relevant.
\EndOfEvidence

So there is plenty of evidence that we do not make much use of evidence.

Reviewing together the examples offered per
respondent finds another disturbing pattern which leads to

%****************************************
\Insight{Itax}{We lack a shared taxonomy of software engineering contexts.}

I will explain this based on how respondents discriminated contexts
when they explained their view of success factors.

\EvidenceI{EtaxA} 
Software engineering settings or contexts are diverse in many respects,
such as team size, development durations, product and release models,
technology used, degrees of reuse, development process models,
quality assurance methods and intensities, software architectures,
and many more.
Most key success factors should not be expected to apply to all of these 
contexts alike.

And yet, some respondents made no context discriminations at all.
Most did, but their discriminations are mostly only binary
and mostly along the same dimension:
low-criticality situations 
(\Genquote{apps})
versus high-criticality situations
(\Genquote{airplanes}).
Cases in between, although relevant, are rarely talked about explicitly.
The context characterization tends to be stereotypical and 
those stereotypes are not accurate.

In particular, most respondents appeared to consider the techniques
applied in high-criticality settings to be of high-tech type,
whereas those few respondents who have actually seen such contexts
characterize them as mostly low-tech, as we already saw above:\\
\rQuote{430}{brute force} or\\
\rQuote{436}{they check and recheck and recheck and
    recheck and recheck.} or\\
\rQuote{415}{And there was a sufficient amount of paranoia. 
  Which you had to have, because you knew that if your software was broken,
  things were going to blow up, people were going to die.
  I don't know how much that actually helped us\Practitioner, though.
  I know we had bugs in that software.}

Besides criticality levels, a number of other dimensions were used to 
discriminate contexts in a manner relevant for the applicability of 
key success factors, but all of those only by one or a few respondents.
The two that appear most relevant are 
when clear expectations about the software's behavior disappear
(\rQuote{472}{I'm very worried as the world embraces machine learning})
or when humans can no longer compensate software misbehavior
(\rQuote{434}{Where the transition, the problems, will happen is as automation 
  increases. There won't be that human backstop that's there to do that 
  last-minute adapation. And I do worry a little bit about that.})
\EndOfEvidence

Summing up, the success factor statements made by the respondents tended to be 
overly general, 
presumably because we have no standard vocabulary in our community 
for thinking and talking about reasonably sized subsectors of our field.

%========================================================================
%========================================================================
\section{Limitations}\label{limitations}

Tracy \cite{Tracy10} suggests a set of eight method-independent quality criteria
for qualitative studies:
(a)~worthy topic, (b)~rich rigor, (c)~sincerity, (d)~credibility,
(e)~resonance, (f)~significant contribution, (g)~ethics, and
(h)~meaningful coherence, each with several facets \cite[Table 1]{Tracy10}.
They are useful for study design by researchers
and for quality assessment by readers.

A credible author self-assessment with these criteria
would require multiple pages,
but in short I see the biggests strengths of the present work
in \emph{(a)~worthy topic} (facets: relevant, significant, interesting)
by the generality of the statements made and the research directions
suggested in Section~\ref{conclusions} and
in \emph{(h)~meaningful coherence} (facets: achieves stated goal,
uses methods that fit the goal, meaningfully interconnects its pieces)
by a gapless train of thought and careful argumentative triangulation
that considers multiple possible views.

The biggest weakness is in \emph{(b)~rich rigor}
(facets: appropriate theoretical constructs, time in the field, samples,
 contexts, analysis processes) for primarily the following reasons:
\begin{itemize}
\item Analysis processes: The deduction rules from Section~\ref{deduction}
  in step 3 rely on inference from my respondents' collective
  interview behavior to the community's knowledge:
  If some knowledge X was not used by any of the
  respondents although it would have been useful for their argumentation,
  I conclude that the community overall is missing that knowledge.
  This conclusion may be invalid if the format of the interviews provided
  insufficient trigger for the respondents to show their knowledge of X.
  It is also invalid if, due to the spontaneous character of the interview,
  the knowledge of X was \emph{momentarily} unavailable to them --
  but if they had had a few days to dig through the literature or simply
  think about what they know, they could have come up with it.
\item Samples: More junior researchers might have added views or evidence
  not currently present in the interview data;
  a few senior respondents I would have liked to have are also
  not present in the data.
  Their presence might have shifted a few of the emphases with respect to
  concepts that had only some or few mentions.
\item Analysis processes: The reduced presentation of the chain of evidence as
  described in Section~\ref{chainofevidence} limits the reader's possibilities
  for checking the analysis underway.
  This decision was made to make the article readable, as a more rigorous
  presentation would have meant to
  (1)~litter the article with many more distracting concept names and
  lengthy concept definitions,
  (2)~do so even for some low-relevance concepts,
  (3)~and even for some first-generation open codes that do not fit into the
  article's train-of-thought at all -- which is why they were replaced
  by the second generation.\\
  Instead, I hope there is sufficient credibility of the presentation
  for most of my readers and refer the others to the raw data package
  for further detail.
\end{itemize}

Of these, the second and third are likely minor, but the first is
a serious threat to the validity of the findings.

%========================================================================
%========================================================================
\section{Related Work}

\subsection{Relating to knowledge gaps}

It appears that newer SE articles talk (in their final section) about
knowledge gaps less often than it was formerly the case:

\EvidenceLit{Enogaptalk}
Out of a random sample of 10 articles from ICSE 1997,
\textbf{5} of them describe knowledge gaps of sorts
\citep{BerBouGal97,JacQue97,KusMizKik97,LinSne97,SeaBas97},
\textbf{3} at least speak of work remaining to be done
\citep{BriDevMel97,ReeLev97,WanRicChe97},
and only \textbf{2} do neither
\citep{FroHooLiu97,VerCer97}.

For ICSE 2018, the corresponding numbers are
\textbf{1} time yes
\citep{HuaZhaZho18},
\textbf{4} times to-do
\citep{CheSuMen18,WanSunChe18,WanSvaWu18,YanSubLu18},
and \textbf{5} times nothing
\citep{BelSprSpi18,HuaZhaWan18,LamStrTae18,WenCheWu18,ZhaLasLyu18}.
The difference from 1997 to 2018 is statistically significant
(Fisher's exact test, $p < 0.01$;
one-sided\footnote{one-sided because I was surprised by the low values
  in 2018 and only then decided to look at 1997, firmly expecting to
  find higher ones.}
asymptotic with-ties Mann-Whitney test, $p < 0.032$).
% i97 = c(1,1,1,1,1, 0.5,0.5,0.5, 0,0);
% i18 = c(1, 0.5,0.5,0.5,0.5, 0,0,0,0,0);
% fisher.test(i97,i18)
% --> p-value = 0.007937
% wilcox.test(i97,i18, alternative='greater')
% --> Wilcoxon rank sum test with continuity correction; p-value = 0.03263
\EndOfEvidence

I interpret this as an increased reluctance to talk about what we do not know.
This is at or near the leaves of the SETT; what about higher up?

Systematic Literature Reviews sometimes conclude there are large gaps
in the knowledge about the respective topic area.
For instance in their SLR on software measurement and process improvement,
Unterkalmsteiner et al.~\cite{UntGorIsl12} conclude 
\emph{``Considering that confounding factors are rarely discussed 
  (19 out of 148 studies [\ldots]), the accuracy of the
  evaluation results can be questioned''},
yet
\emph{``no good conceptual model or framework for such a discussion is currently
  available.''}
And Hannay et al.~\cite{HanDybAri09},
after their quantitative meta-analysis of pair programming efficiency,
conclude that knowledge about more complex, qualitative
factors is low:
\emph{``Only by understanding what makes pairs work, and what makes pairs 
  less efficient, can steps be taken to provide beneficial conditions for
  work and to avoid detrimental conditions''}.
So much for medium levels of the SETT; what about still higher up?

This brings us back to the discussion from Section~\ref{stateofknowledge}.
More pertinent work, if it exists, is difficult to find, because all
obvious search terms for it are impractically broad.
Several of the sources in the following subsection \emph{could} be pertinent
(and a few are, to a small degree),
but they all focus on knowledge, not knowledge gaps.

\subsection{Relating to the interview content}

Is there literature that answers QQ?
The literature most related to QQ is that on critical success factors.
There are such works 
for domains only partly related to SE, such as 
general project management \citep{Cooke02}
or information systems \citep{NgaLawWat08}.
Within SE, there are works for subdomains,
such as agile methods \citep{ChoCao08},
or using specific perspectives, such as
CMMI \citep{GolGib03} or
project management failures \citep{Ewusi03}.

If we count out textbooks,
there appear to be only few works with a global perspective.
Hoare \cite{Hoare96} asks
\emph{``How did software get so reliable without proof?''}.
This looks pertinent, but
turns out to be a position paper with no evidence and not even references.
The Handbook of Endres and Rombach~\cite{EndRom03}
is focused on references to evidence, but is unopinionated with 
respect to discriminating key factors from less central ones.
The latter is also true for the SWEBOK Guide \citep{SWEBOK14},
which furthermore does not emphasize evidence.

The best match appears to be Johnson et al.~\cite{JuhRalEks18},
which asked a pre-structured form of QQ in 2013:
They derive 28 SE success factors and ask each of 60 ICSE participants to
arrange his or her personal top-10 of those factors 
into a boxes-and-arrows diagram
of how they impact ``software engineering success''.
The authors cluster these diagrams into three recurring types and point out
that finding multiple types implies a lack of consensus in our field.
SE context discriminations are actively
\emph{suppressed} by the study design \cite[Section 2.1]{JuhRalEks18}.
Also, the factors mentioned are at least as vague as in the present study,
but this is not discussed at all.

With respect to SE community knowledge gaps,
all four sources mention that gaps exist, but none of them work out
which gaps appear to have key roles for the state of our knowledge.

A diverse set of literature was suggested by my respondents (some of
it after the interview when I asked by email),
e.g. about S1-not-as-high-as-one-might-think complexity
\citep{AllBarDev19,MenOweRic07,PetHarHar18},
% !!!consider Barr: Bimodal Software Engineering if it appeared.
% The post-deployment usability articles are insufficiently on-topic to be used
about the possibility of catastrophes due to our limited understanding
\cite[regarding S1, S2, S3, QQ]{Leveson92},
about Lehman's astonishingly fresh and comprehensive laws of software evolution
\cite[regarding S1, S2, S3, QQ]{Lehman80},
about good-enoughness
\cite[regarding S2, QQ]{Shaw02},
about the role of developers as people and as knowledge carriers beside
and beyond the program code
\cite[regarding S1, QQ]{Naur85},
about research with industrial relevance
\cite[regarding IA]{BasBriBia19},
or about the astonishing amount of knowledge and collaboration required
for producing something as simple as a pencil
\cite[regarding QQ]{Read58}.

Also, Dijkstra's Turing Lecture \cite{Dijkstra72}, 
when read from an S1/S3/QQ/IA perspective,
is not only as keen as we would expect, but also surprisingly fresh.
What's missing is mostly the social dimension, on both the development
and the usage side.

%========================================================================
%========================================================================
\section{Conclusions and Further Work}\label{conclusions}

\subsection{Insights}

We (the SE research community) are concerned a lot with complexity,
but hardly understand how it comes to be (see Section~\ref{S1}).
We are aware that perfection is not typically to be achieved,
but barely understand how quality tradeoffs are made (see Section~\ref{S2}).
We research or invent techniques without giving much consideration
to the conditions under which they will or will not be helpful
(see Section~\ref{QQ}).

\subsection{ETAT: Suggested research directions}

The above points suggest a number of research directions that should 
presumably be emphasized more in SE research.

\Heading{1. Focus on emergence}
(see Insight~\ref{Icplx}):
Assuming the indirect complexity effects are harder to understand and handle
than direct ones, we should do more research to understand
where and how complexity phenomena emerge from constituent parts and events.

\Heading{2. Focus on tradeoffs}
(see Insight~\ref{Igoodenough}):
Assuming that good-enoughness and developer limitations have near-ubiquitous
relevance for impactful SE research, 
we should do more research to understand
where, how, and how well engineers make tradeoffs 
in practical SE situations.

\Heading{3. Evaluate assumptions}
(see Insight~\ref{Itax}):
Assuming that differences between software engineering contexts are
relevant,
we should collect and tabulate all the assumptions, tacit or explicit,
that are used for justifying and scoping SE research
in order to determine which different contexts might be relevant.
These assumptions should then be evaluated:
When are they true?
When are they not-so-true?
To which degree?

\Heading{4. Create taxonomies}
(see Insight~\ref{Iebse}):
Based on these results we could then define community-wide terminology
that allows to state assumptions and claims more precisely than we do today,
in order to become a properly evidence-based discipline.
Such terminology would define taxonomies of
the different sources and kinds of complexity (from emergence analysis),
the different natures of engineering decisions (from tradeoffs analysis), and
the different contexts in which we attempt to produce software 
(from assumptions analysis and field work).

For refering to these goals together, I suggest we call
them \textbf{ETAT} goals (acronym of emergence, tradeoffs, assumptions,
taxonomies; also French for \emph{state} or \emph{constitution} or 
various other things).

\subsection{Next step}

TSE reviewers of this work found the problem with the deduction rule
(as discussed in Section \ref{limitations}) so big that they insisted
the insights relying on the rule need to be independently validated.

So a survey asking about those statements and about possible counter-examples
of work providing the respective knowledge is the next step in this research.

%========================================================================
%========================================================================
\section*{Acknowledgments}

I dearly thank my interviewees for their time and explanations.
I thank Franz Zieris and Kelvin Glaß for sharing their views 
as pilot respondents.
I am indebted to Franz Zieris for his excellent contribution as
advocatus diaboli.

% set BIBINPUTS=../bib or the like
\bibliographystyle{IEEEtran}
\bibliography{sehow}

\begin{IEEEbiography}[{%
\includegraphics[width=1in,height=1.25in,clip,keepaspectratio]%
{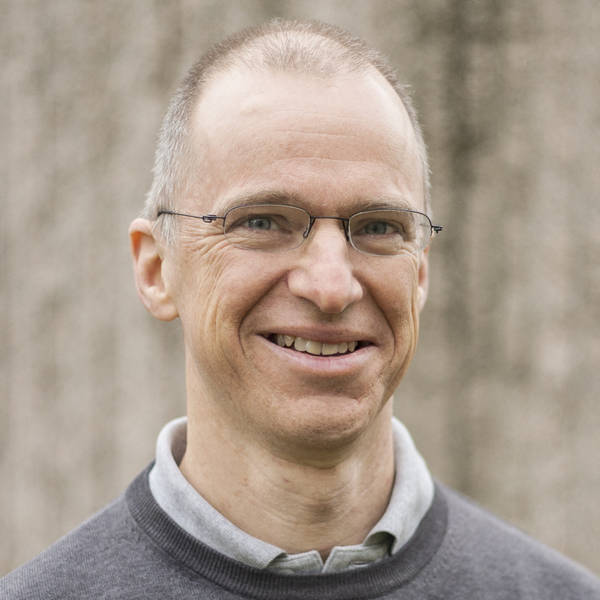}}]{Lutz Prechelt}
Lutz Prechelt received a PhD from the University of Karlsruhe
for work that combined machine learning and compiler construction
for parallel machines.
He then moved to empirical software engineering and performed
a number of controlled experiments before spending three
years in industry as a manager.
He is now full professor for software engineering at
Freie Universität Berlin.
His research interests concern the human
factor in the software development process, asking mostly
exploratory research questions and addressing them with
qualitative methods.
Additional research interests concern research methods
and the health of the research system.
He is the founder of the Forum for Negative Results
and of Review Quality Collector.
\end{IEEEbiography}

\end{document}